\documentclass[conference]{IEEEtran}

\usepackage[utf8]{inputenc}
\usepackage[T1]{fontenc}
\usepackage[english]{babel} 

\usepackage{cite}
\usepackage{amsfonts,amsmath,amssymb}
\usepackage{array}
\usepackage{subfigure}
\usepackage{url}
\usepackage{siunitx}
\DeclareSIUnit\rpm{rpm}

\usepackage{ae}
\usepackage{aecompl}
\usepackage{aeguill}

\usepackage{graphicx} 

\usepackage{booktabs}

\graphicspath{{Figures/}}

\usepackage{tikz}
\usetikzlibrary{calc}
\usepackage{pgfplots}
\pgfplotsset{compat=newest}
\pgfplotsset{plot coordinates/math parser=false}
\tikzstyle{dotted} = [line width = 0.5pt, dash pattern = on \pgflinewidth off 2\pgflinewidth]
\pgfplotsset{%
	tick label style={font=\scriptsize,%
		/pgf/number format/precision=3,%
		/pgf/number format/std=-3:3},%
	legend style={font=\small},%
	label style={font=\small},%
	every axis post/.append style={label style={font=\small},%
		axis line style=-,%
		scaled ticks=false}%
}

\newlength{\figurewidth}
\newlength{\figureheight}

\def\ordfield{}
\def\subfield{}
\def\expfield{}
\def\getfields#1{%
	\def\ordfield{}%
	\def\subfield{}%
	\def\expfield{}%
	\expandafter\getexpfield\expandafter\beginsym#1^!\endsymb%
}
\def\extractsubfield\beginsymb#1_!\endsymb{%
	\def\subfield{#1}%
}
\def\getsubfield\beginsymb#1_#2\endsymb{%
	\if!#2\relax\else%
		\extractsubfield\beginsymb#2\endsymb%
	\fi%
	\def\ordfield{#1}%
}
\def\extractexpfield\beginsymb#1^!\endsymb{%
	\getsubfield\beginsymb#1_!\endsymb%
	\expandafter\def\expandafter\expfield\expandafter{\ordfield}
}
\def\getexpfield\beginsym#1^#2\endsymb{%
	\if!#2\relax\else%
		\extractexpfield\beginsymb#2\endsymb%
	\fi%
	\getsubfield\beginsymb#1_!\endsymb%
}

\newcommand{\be}{\begin{equation}}
\newcommand{\ee}{\end{equation}}
\newcommand{\beno}{\begin{equation*}}
\newcommand{\eeno}{\end{equation*}}
\newcommand{\barCl}{\begin{IEEEeqnarray}{rCl}}
\newcommand{\earCl}{\end{IEEEeqnarray}}
\newcommand{\barClno}{\begin{IEEEeqnarray*}{rCl}}
\newcommand{\earClno}{\end{IEEEeqnarray*}}

\newcommand{\AB}{{\alpha\beta}}		
\newcommand{\DQ}{{DQ}}				
\newcommand{\HmDQ}{\mathcal{H}^\DQ}	
\newcommand{\HMDQ}{\HmDQ_m}			
\newcommand{\us}{u_s}
\newcommand{\usDQ}{\us^\DQ}			
\newcommand{\isDQ}{{\imath_s^\DQ}}	
\newcommand{\isAB}{{\imath_s^\AB}}	
\newcommand{\isD}{{\imath_s^D}}		
\newcommand{\isQ}{{\imath_s^Q}}		
\newcommand{\PhisDQ}{{\Phi_s^\DQ}}	
\newcommand{\PhisD}{{\Phi_s^D}}	
\newcommand{\PhisQ}{{\Phi_s^Q}}	
\newcommand{\phisDQ}{{\phi_s^\DQ}}	
\newcommand{\phisD}{{\phi_s^D}}		
\newcommand{\phisQ}{{\phi_s^Q}}		
\newcommand{\Te}{T_e}				
\newcommand{\Tl}{T_l}				
\newcommand{\we}{\omega}			
\newcommand{\te}{\theta}			

\newcommand{\LD}{L^D}				
\newcommand{\LQ}{L^Q}				
\newcommand{\PhiM}{\Phi_M}			
\newcommand{\Rs}{R_s}				
\newcommand{\Jl}{J_l}				
\newcommand{\np}{n}					
\newcommand{\W}{\Omega}				
\newcommand{\finj}{f}
\newcommand{\Finj}{F}
\newcommand{\foinj}{f_i}
\newcommand{\Sal}{\mathcal{S}}		
\newcommand{\JJ}{\mathcal{J}}		
\newcommand{\Rot}{\mathcal{R}}		

\newcommand{\transpose}[1]{{#1}^T}

\newcommand{\ppderive}[2]{\partial^2_{#2}#1}

\newcommand{\pderive}[2]{\partial_{#2}#1}

\newcommand{\Tderive}[2][1]{%
	\if1#1\relax
		\frac{d#2}{dt}%
	\else%
		\frac{d^{#1}#2}{dt^{#1}}%
	\fi%
}

\DeclareMathOperator{\Landau}{\mathcal{O}}

\newcommand{\lf}[1]{\overline{#1}}
\newcommand{\hf}[1]{\widetilde{#1}}

\newcommand{\telr}{\te_{lr}}			
\newcommand{\jsDQ}{{I_s^\DQ}}	
\newcommand{\jsD}{{I_s^D}}		
\newcommand{\jsQ}{{I_s^Q}}		
\newcommand{\gAB}{\gamma^\AB}	

\newcommand{\vs}{v_s}
\newcommand{\vsDQ}{\vs^\DQ}			
\newcommand{\vsAB}{\vs^\AB}			
\begin{document}
	%
	\title{Obtaining the Current-Flux Relations of the Saturated PMSM by Signal Injection}

%
	
	%
	\author{\IEEEauthorblockN{Pascal Combes\IEEEauthorrefmark{1},
	François Malrait\IEEEauthorrefmark{1},
	Philippe Martin\IEEEauthorrefmark{2} and
	Pierre Rouchon\IEEEauthorrefmark{2}}
	\IEEEauthorblockA{\IEEEauthorrefmark{1}Schneider Toshiba Inverter Europe, 27120~Pacy-sur-Eure, France\\
	Email: \{pascal.combes, francois.malrait\}@schneider-electric.com}
	\IEEEauthorblockA{\IEEEauthorrefmark{2}Centre Automatique et Systèmes, MINES ParisTech, PSL Research University, 75006 Paris, France\\
	Email: \{philippe.martin, pierre.rouchon\}@mines-paristech.fr}}
	

	\maketitle
	
	\begin{abstract}
			This paper proposes a method based on signal injection to obtain the saturated current-flux relations of a PMSM from locked-rotor experiments. With respect to the classical method based on time integration, it has the main advantage of being completely independent of the stator resistance; moreover, it is less sensitive to voltage biases due to the power inverter, as the injected signal may be fairly large. 
	\end{abstract}
	

	%
	\IEEEpeerreviewmaketitle

\section{Introduction}\label{sec:intro}
Good models are usually paramount to design good control laws. This is the case for Permanent Magnet Synchronous Motors (PMSM), especially when ``sensorless'' control is considered. In this mode of operation, neither the rotor position nor its velocity is measured, and the control law must make do with only current measurements; a suitable model is therefore essential to relate the currents to the other variables. When operating above moderately low speed, i.e., above about $10\%$ of the rated speed, models neglecting magnetic saturation are usually accurate enough for control purposes; but at low speed, magnetic (cross-)saturation must be taken into account, in particular when high-frequency signal injection is used, and the more so for motors with little geometric saliency, see e.g.~\cite{ReigoGR2008ITIA,LiZHBS2009ITIA,SergeDM2009ITM,BiancFB2013ITIA,JangSHIS2003ITIA,JebaiMMR2016IJC}.

However, motor manufacturers very seldom provide saturation data, which means the saturated current-flux relations must be experimentally determined. The ``classical'' method to get these data is to integrate over time the time derivative of the flux as given by the stator model of the motor, see~\cite{StumbPSTD2005ITM} for a detailed account. Unfortunately, this method is very sensitive to the value of the stator resistance~$\Rs$, which may significantly vary during a long experiment; it also requires a good knowledge of the actually impressed potentials (voltage sensors may be required because of the not very well-known voltage drops in the power stage), and of course of the resulting currents (which are in practice always measured). It is thus not easy to implement this method on a commercial variable speed drive for use in the field.

The goal of this paper is to propose an alternative approach to obtain the saturated current-flux relations, which is completely independent on the knowledge of the stator resistance~$\Rs$; it is also less sensitive to the voltage drops of the power stage. It is based on high-frequency signal injection. It is based on signal injection, a technique that was originally introduced for sensorless control at low velocity~\cite{JanseL1995ITIA}, but that can also be used for identification~\cite{JebaiMMR2011IEMDC}. The key idea, thanks to a suitable analysis of the effects of signal injection, is to recover the flux-current relations by integrating over paths in the plane of direct and quadrature currents; it generalizes a cruder procedure used in~\cite{CombeMMR2011IEMDC}.
Notice that the required experimental data can be obtained with only experiments where the rotor is locked in a known position, which is reasonable for industrial use in the field (the classical method also works in similar conditions).

The paper runs as follows: section~\ref{sec:model} presents the structure of the model, based on an energy approach; section~\ref{sec:classical} applies the classical method to a test PMSM; section~\ref{sec:proposed} details the proposed approach, and applies it to the same test PMSM.

The following conventions are used: if $x^{ij}$, $ij$ being $\AB$ or $\DQ$, is the vector with coordinates~$x^i$ and $x^j$, we write indifferently $x^{ij}$ and $(x^i,x^j)^T$; if $f$ is a function of several variables, $\pderive{f}{k}$ denotes its partial derivative with respect to the $k^{th}$ variable, and $\ppderive{f}{kl}=\ppderive{f}{lk}$ the second partial derivative with respect to the $k^{th}$ and $l^{th}$ variables.
Lastly, the rotation matrix of angle~$\alpha$ is denoted by {\footnotesize$\Rot(\alpha):=\begin{pmatrix}
\cos{\alpha} & -\sin{\alpha}\\
\sin{\alpha} & \cos{\alpha}
\end{pmatrix}$};
of course, {\footnotesize$\Rot(\alpha)\transpose\Rot(\alpha)=\begin{pmatrix}1 & 0\\ 0 & 1\end{pmatrix}$}, where $\transpose\Rot(\alpha)$ is the transpose of~$\Rot(\alpha)$.

\section{Energy-based model of the PMSM}\label{sec:model}
To write a model of the star-connected saturated sinusoidal PMSM in the $\DQ$~frame, we follow the energy-based approach of~\cite{JebaiCMMR2014CDC,CombeJMMR2016arXiv}. All the motor specific information is encoded in the scalar magnetic energy function $\HMDQ(\phisDQ)$, where $\phisDQ$ is the flux linkage vector. $\HMDQ$ is independent of the (electrical) rotor angle~$\te$ by the assumption of sinusoidal windings, and independent of the (electrical) rotor velocity~$\we$ as in any conventional electromechanical device. The state equations of the PMSM then read
\begin{IEEEeqnarray}{rCl}\label{eqn:model:dynamic}
	\Tderive{\phisDQ} &=& \vsDQ - \Rs\isDQ - \we\JJ\phisDQ\label{eqn:model:dynamic:phi}\\
	\frac{\Jl}{\np} \Tderive{\we} &=& - \np\transpose{\phisDQ}\JJ\isDQ - \Tl\label{eqn:model:dynamic:w}\\
	\Tderive{\te} &=& \we,\label{eqn:model:dynamic:w}
\end{IEEEeqnarray}
where $\vsDQ$ is the impressed potential vector, $\Rs$ the stator resistance, $\np$ the number of pole pairs, $\Tl$ the load torque and $\footnotesize\JJ:=\begin{pmatrix}0 & -1\\ 1 & 0\end{pmatrix}$; the stator current vector~$\isDQ$ is the gradient of $\HMDQ$, i.e.,
\begin{IEEEeqnarray}{rCl}\label{eqn:model:algebraic}
\isDQ &:=& \nabla\HMDQ(\phisDQ)=
\begin{pmatrix}
	\pderive{\HMDQ}{1}(\phisD,\phisQ)\\ \pderive{\HMDQ}{2}(\phisD,\phisQ)
\end{pmatrix}.
\end{IEEEeqnarray}
The physical control input is the potential vector $\vsAB:=\Rot(\te)\vsDQ$ impressed in the $\AB$~frame, i.e., the (fictitious) potential vector $\vsDQ$ rotated by~$\Rot(\te)$.
Similarly, the measured current vector is $\isAB:=\Rot(\te)\isDQ$; in sensorless control, this is the only available measurement.

Taking advantage of the construction symmetries of the PMSM (the stator and the rotor of the PMSM are symmetric with respect to a plane), see~\cite{CombeJMMR2016arXiv} for details, we can moreover write
\begin{IEEEeqnarray}{rCl}\label{eqn:model:symmetries}
	\HMDQ(\phisD, -\phisQ) &=& \HMDQ(\phisD,\phisQ),
\end{IEEEeqnarray}
i.e., the magnetic energy function~$\HMDQ$ of a PMSM is even with respect to the $q$-axis flux linkage~$\phisQ$. 

Notice that thanks to the assumption of sinusoidal windings, all that is needed to close the model~\eqref{eqn:model:dynamic:phi}--\eqref{eqn:model:dynamic:w} are the flux-current relations~\eqref{eqn:model:algebraic}; this is no longer the case for non-sinusoidal windings, since the electro-magnetic torque in~\eqref{eqn:model:dynamic:w} explicitly depends on~$\HMDQ$.
Notice also that the simplest acceptable function is the quadratic form
\begin{IEEEeqnarray*}{rCl}
	\HMDQ(\phisDQ) &:=& \frac{1}{2\LD}(\phisD-\PhiM)^2 + \frac{1}{2\LQ}\phisQ^2,
\end{IEEEeqnarray*}
which yields the current-flux relations
\begin{IEEEeqnarray*}{rCl}
	\isD &=& \frac{\phisD-\PhiM}{\LD}\\
	\isQ &=& \frac{\phisQ}{\LQ}
\end{IEEEeqnarray*}
and the electro-magnetic torque
\begin{IEEEeqnarray*}{rCl}
\Te &=& \frac{\np}{\LD}\PhiM\phisQ + \np\biggl(\frac{1}{\LQ} - \frac{1}{\LD}\biggr)\phisD\phisQ;
\end{IEEEeqnarray*}
in other words, the simplest acceptable magnetic energy function represents the unsaturated PMSM. The unsaturated model is usually sufficient for control above moderately low speed.

This energy approach enjoys several interesting features:
\begin{itemize}
	\item it naturally enforces the reciprocity conditions $\frac{\partial\isD}{\partial\phisQ}=\frac{\partial\isQ}{\partial\phisD}$  \cite{MelkeW1990IAITo}, since $\ppderive{\HMDQ}{12}(\phisD,\phisQ)=\ppderive{\HMDQ}{21}(\phisD,\phisQ)$
	\item it yields a valid expression for the magnetic torque, even in the presence of magnetic saturation
	\item it justifies the modeling of saturation in the fictitious rotor $\DQ$~frame for a star-connected motor. Though this point is usually taken for granted, it is not completely obvious because of the nonlinearities due to saturation that: i) the transformation from the physical $abc$-frame to the fictitious $\DQ0$-frame behaves well; ii) the decoupling between the $\DQ$- and the 0-axes is still valid
	\item it requires only a very basic knowledge of the motor internal layout
	\item finally, it is particularly amenable to an analysis of the effects of signal injection.
\end{itemize}

In all the experiments, the rotor will be locked in a known position, so that~\eqref{eqn:model:dynamic:phi}--\eqref{eqn:model:dynamic:w} reduces to
\begin{IEEEeqnarray}{rCl}\label{eqn:model:locked}
	\Tderive{\phisDQ} &=& \transpose\Rot(\telr)\vsAB - \Rs\isDQ,
\end{IEEEeqnarray}
with $\telr$ constant and known. Notice that since $\telr$ is known, we can consider that $\vsDQ:=\transpose\Rot(\telr)\vsAB$ is the impressed potential, and $\isDQ:=\transpose\Rot(\telr)\isAB$ is the available measurement.

\section{Classical method}\label{sec:classical}
The most widely used method to obtain the current-flux relations, see~\cite{StumbPSTD2005ITM}, assumes that an impressed potential trajectory $t\mapsto\vsDQ(t)$ is known, together with the resulting current trajectory $t\mapsto\isDQ(t)$. The corresponding flux linkage is obtained by time integrating~\eqref{eqn:model:locked}, i.e.,
\begin{IEEEeqnarray}{rCl}\label{eq:timeIntegration}
	\phisDQ(t) &=& \phisDQ(0)+\int_0^t\bigl(\vsDQ(\tau) - \Rs\isDQ(\tau)\bigr) d\tau,
\end{IEEEeqnarray}
the unknown initial value $\phisDQ(0)$ being yet to determine. A pair $\bigl(\isDQ(t),\phisDQ(t)\bigr)$ is thus obtained for each time~$t$; provided the current trajectory $t\mapsto\isDQ(t)$ covers a sufficient area of the current plane, this yields the desired current-flux relations. The initial value $\phisDQ(0)$ is chosen so as the flux linkage is zero when the current is zero.

Thanks to the filtering effect of the integration, the method is rather insensitive to measurement noise. However, it is strongly affected by biases in:
\begin{itemize}
	\item the impressed potential. The power stage is usually an IGBT bridge commuted with PWM; owing to voltage drops in the transistors and dead times, the actually impressed potential somewhat differs from the desired one, see e.g.~\cite{WeberS2012IIIMTCP}. This is not a problem if voltage sensors are available, as in a laboratory experiment, but matters for implementation on industrial drives, which are usually not equipped with such sensors 
	\item the stator resistance~$\Rs$ estimation; this is the main problem, since the value of the resistance can significantly vary during a long experiment.
\end{itemize}

\subsection{Experimental results}\label{sec:classical:xp}
\begin{figure}
	\centering
	\includegraphics{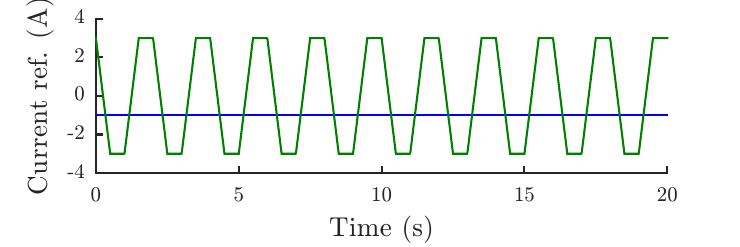}\\
	\includegraphics{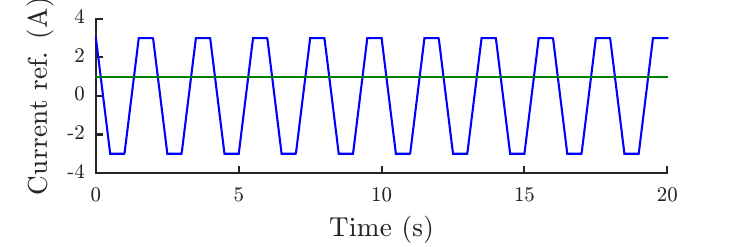}\\
	\includegraphics{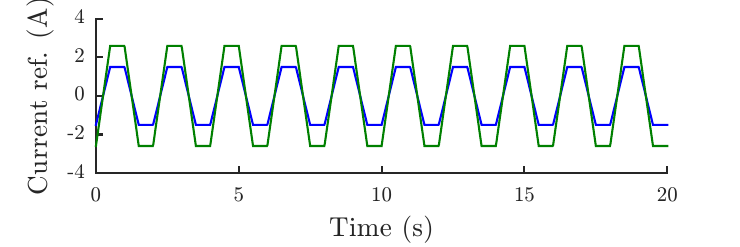}
	\caption{The three current trajectories on $D$- (\textcolor{blue}{---}) and $Q$-axis (\textcolor{green!50!black}{---}) used in the experiment.}
	\label{fig:classicalxp:current-ref}
\end{figure}

The method was used to obtain the current-flux curves of a \SI{400}{\watt} PMSM (rated parameters in table~\ref{tbl:classicalxp:motor}). The motor is fed by an \SI{1.5}{\kilo\watt}~ATV71 (ATV71HU15N4) inverter bridge driven by a dSpace board (DS1005).
The rotor was first aligned so that the $\AB$ and $\DQ$~frames coincide, and then locked with a mechanical brake.
%
\begin{table}
	\centering
	\caption{Rated parameters of test PMSM.}
	\[\begin{array}{lr}
	\toprule
	\mbox{Rated power} & \SI{400}{\watt}\\
	\mbox{Rated voltage (RMS)} & \SI{139.3}{\volt}\\
	\mbox{Rated current (RMS)} & \SI{1.66}{\ampere}\\
	\mbox{Rated frequency} & \SI{60}{\hertz}\\
	\mbox{Rated speed} & \SI{1800}{\rpm}\\
	\mbox{Rated torque} & \SI{2.12}{\newton\meter}\\
	\midrule
	\mbox{Number of pole pairs $\np$} & \si{2}\\
	\mbox{Stator resistance $\Rs$} & \SI{4.25}{\ohm}\\
	\mbox{$D$-axis inductance $\LD$} & \SI{43.25}{\milli\henry}\\
	\mbox{$Q$-axis inductance $\LQ$} & \SI{69.05}{\milli\henry}\\
	\bottomrule
	\end{array}\]
	\label{tbl:classicalxp:motor}
\end{table}
To check the consistency of the results, three trajectories were used:
\begin{itemize}
	\item constant $\isD$, \SI{2}{\second}-periodic trapezoidal $\isQ$ with amplitude~\SI{3}{\ampere}, see fig.~\ref{fig:classicalxp:current-ref}(top); the experiment was repeated for $\isD:=\text{\SIlist{-3;-2;-1;0;1;2;3}{\ampere}}$
	\item constant $\isQ$, \SI{2}{\second}-periodic trapezoidal $\isD$ with amplitude~\SI{3}{\ampere}, see fig.~\ref{fig:classicalxp:current-ref}(middle); the experiment was repeated for $\isQ:=\text{\SIlist{-3;-2;-1;0;1;2;3}{\ampere}}$
	\item proportional \SI{2}{\second}-periodic trapezoidal currents on both axes with total amplitude~\SI{3}{\ampere}, see fig.~\ref{fig:classicalxp:current-ref}(bottom).
\end{itemize}
To enforce these trajectories, the currents were controlled with Proportional-Integral controllers on the $D$- and $Q$-axes (damping ratio $\xi:=\frac{1}{\sqrt2}$; bandwidth~$\omega_0:=\SI{25}{\hertz}$).

The stator resistance is evaluated by computing the ratio between the voltage and the current during the phases when the current is constant; it is found to vary from \SIrange{4.5}{5.25}{\ampere} over the whole experiment.

\begin{figure}
	\centering
	\includegraphics{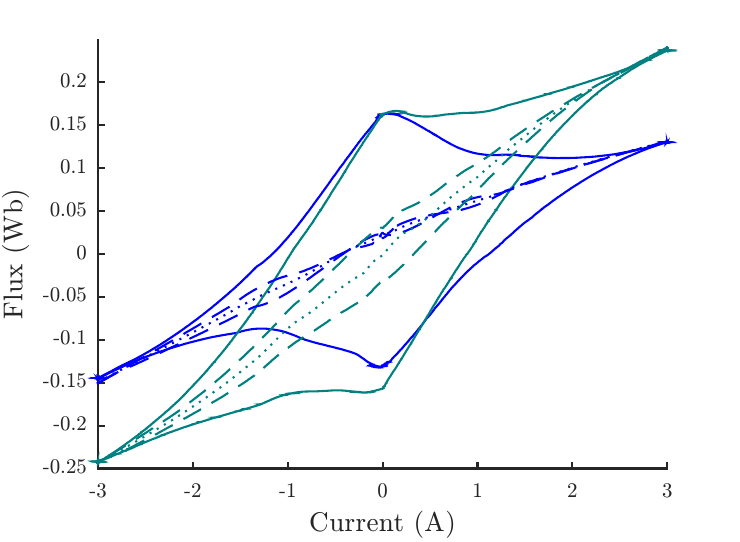}
	\caption{Experimental current-flux curves $\phisD(\isD,0)-\PhiM$ (\textcolor{blue}{---}) and $\phisQ(0,\isQ)$ (\textcolor{green!50!black}{---}); solid lines: uncompensated voltage drops; dashed lines: compensated voltage drops; dotted line: averaged curve.}
	\label{fig:classicalxp:processing}
\end{figure}

The flux linkage is then computed according to~\eqref{eq:timeIntegration}. The sensitivity to voltage biases is illustrated in fig.~\ref{fig:classicalxp:processing}: when the time integration is performed over several identical similar patterns (10 in our case), the shape of the experimental current-flux curve is altered, especially when the current is small; with uncompensated inverter voltage drops and dead times, the result is very bad (solid lines); with compensation by a suitable model, the loop in the current-flux curve is much smaller, but still present (dashed lines);
the final current-flux curve is then obtained by averaging all the flux linkages points obtained for a current point (dotted line).

\begin{figure}
	\centering
	\subfigure[$D$-axis flux linkage in function of $\isDQ$\label{fig:classicalxp:results:d}]{\includegraphics{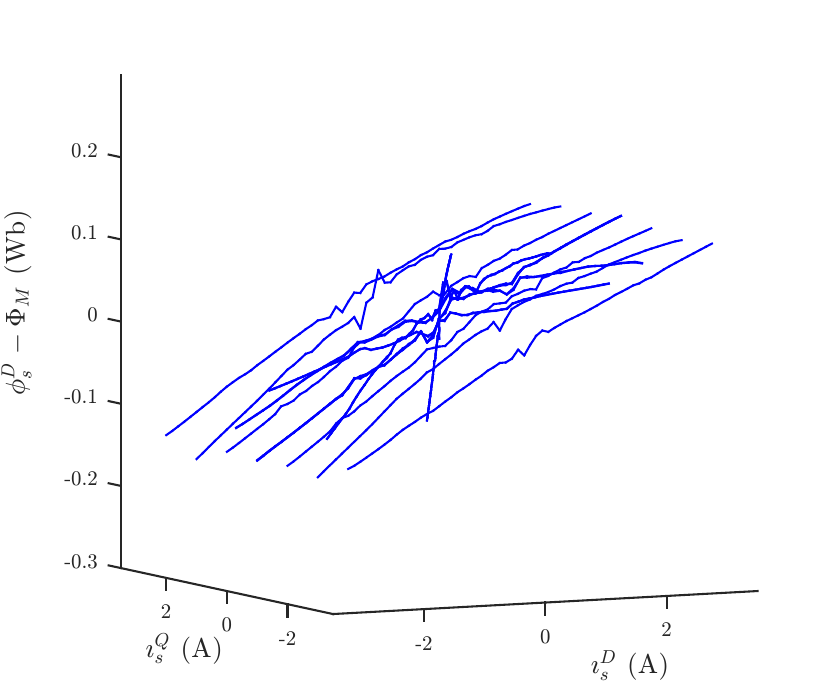}}
	\subfigure[$Q$-axis flux linkage in function of $\isDQ$\label{fig:classicalxp:results:q}]{\includegraphics{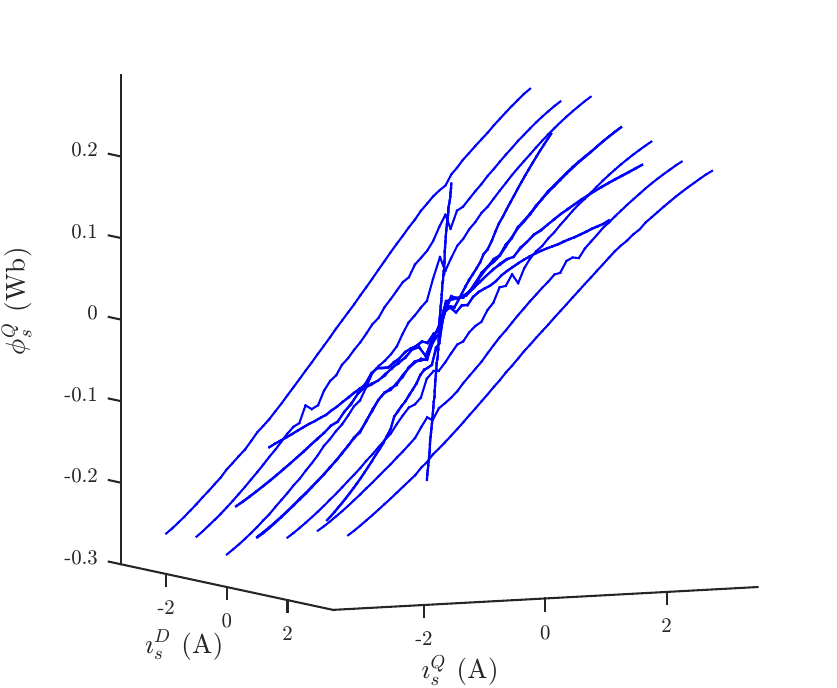}}
	\caption{Current-flux relations obtained by the classical method.}
	\label{fig:classicalxp:results}
\end{figure}

The full 3D current-flux relations are displayed in fig.~\ref{fig:classicalxp:results}.
The magnetic saturation is clearly visible. As expected from~\eqref{eqn:model:symmetries}, which implies
\begin{IEEEeqnarray*}{rCl}
	\pderive{\HMDQ}{1}(\phisD,-\phisQ) &=& \pderive{\HMDQ}{1}(\phisD,\phisQ)\\ 
	\pderive{\HMDQ}{2}(\phisD,-\phisQ) &=& -\pderive{\HMDQ}{1}(\phisD,\phisQ), 
\end{IEEEeqnarray*}
it can be seen that $\phisD$ is even with respect to $\isQ$, and $\phisQ$ odd with respect to $\isQ$.

\section{Proposed method}\label{sec:proposed}
\subsection{Data acquisition with signal injection}\label{sec:proposed:hf}
Signal injection was originally introduced for sensorless control at low velocity~\cite{JanseL1995ITIA}, but it can also be used for identification~\cite{JebaiMMR2011IEMDC}. We follow the quantitative analysis introduced in~\cite{JebaiMMR2011IEMDC,JebaiMMR2016IJC} and studied in detail in~\cite{CombeJMMR2016ACC}; it is valid even in the presence of nonlinearities due to magnetic saturation, and whatever the shape of the (periodic) injected signal. Impressing in~\eqref{eqn:model:locked} a potential vector of the form
\be\label{eqn:hf:injection}
\vsAB = \lf{\vsAB} + \hf{\vsAB}\finj(\W t),
\ee
where $\finj$ a 1-periodic function and $\W$~is a ``large'' frequency, the analysis shows that the actual flux linkage is
\begin{IEEEeqnarray}{rCl}
	\phisDQ &=& \lf{\phisDQ} + \frac{1}{\W}\transpose\Rot(\telr)\hf{\vsAB} \Finj(\W t) + \Landau\bigl(\tfrac{1}{\W^2}\bigr),\label{eqn:hf:phi}
\end{IEEEeqnarray}
where $\Finj$~is the primitive of~$\finj$ with zero mean, and $\Landau$ is the ``big-O'' symbol of analysis; $\lf{\phisDQ}$ is the flux linkage without signal injection, i.e., the solution of
\begin{IEEEeqnarray*}{rCl}
	\Tderive{\lf{\phisDQ}} &=& \transpose\Rot(\telr)\lf{\vsAB} - \Rs\lf{\isDQ}.
\end{IEEEeqnarray*}
Plugging~\eqref{eqn:hf:phi} into~\eqref{eqn:model:algebraic} and expanding then yields
\begin{IEEEeqnarray}{rCl}\label{eqn:hf:mes}
	\isAB &=& \lf{\isAB} + \frac{1}{\W}\hf{\gAB} \Finj(\W t) + \Landau\bigl(\tfrac{1}{\W^2}\bigr),
\end{IEEEeqnarray}
where $\hf{\gAB}:=\Sal\bigl(\telr,\lf{\phisDQ}\bigr)\hf{\vsAB}$ and
\begin{IEEEeqnarray*}{rCl}\label{eqn:hf:sal}
	\Sal(\theta,\phi)
	&:=& \Rot(\theta)
	\begin{pmatrix}
		\ppderive{\HMDQ}{11}(\phi) & \ppderive{\HMDQ}{12}(\phi)\\
		\ppderive{\HMDQ}{12}(\phi) & \ppderive{\HMDQ}{22}(\phi)	
	\end{pmatrix}\transpose\Rot(\theta).
\end{IEEEeqnarray*}
We call $\Sal\bigl(\telr,\lf{\phisDQ}\bigr)$ the ``saliency matrix''; indeed, it effectively depends on~$\telr$ if the motor exhibits saliency, whether geometric or induced by magnetic saturation.
In other words, \eqref{eqn:hf:mes} shows that a small ripple of amplitude $\hf{\isAB}:=\frac{1}{\W}\hf{\gAB}$, produced by the injected signal~$\hf{\vsAB}\finj(\W t)$, is superimposed to the current without signal injection~$\lf{\isAB}$, produced by~$\lf{\vsAB}$. As explained in~\cite{CombeJMMR2016ACC}, both $\lf{\isAB}$ and~$\hf{\isAB}$ can be extracted from the actual measurement~$\isAB$ using the estimations
\begin{IEEEeqnarray*}{rCl}
	\lf{\isAB}(t) &=& \W \int_{t-\frac{1}{\W}}^{t} \isAB(\tau) d\tau \\
	\hf{\isAB}(t) &=& \W \frac{\int_{t-\frac{1}{\W}}^{t} \Bigl(\isAB\bigl(\tau - \frac{1}{2\W}\bigr) - \lf{\isAB}(\tau)\Bigr) \Finj(\W\tau) d\tau}{\int_{t-\frac{1}{\W}}^{t} \Finj^2(\W t) dt}.
\end{IEEEeqnarray*}
In other words, the ``virtual'' measurement $\hf{\isAB}$ has been made available besides the ``actual'' measurement~$\lf{\isAB}$.

Notice that since $\telr$ is known, we can consider that $\lf{\vsDQ}:=\transpose\Rot(\telr)\lf{\vsAB}$ and $\hf{\vsDQ}:=\transpose\Rot(\telr)\hf{\vsAB}$ are the impressed potentials, and that $\lf{\isDQ}:=\transpose\Rot(\telr)\lf{\isAB}$ and $\hf{\isDQ}:=\transpose\Rot(\telr)\hf{\isAB}$, are the available measurements. It is then possible to experimentally acquire the six expressions

Experimentally, it is possible to apply any desired current $\lf{\isDQ}$ by a suitable choice of $\lf{\vsDQ}$; by signal injection with at least two independent vectors $\hf{\vsDQ}$ and extraction of the corresponding~$\hf{\isDQ}$, it is then possible to obtain the complete saliency matrix, hence $\ppderive{\HMDQ}{11}(\lf{\phisDQ})$, $\ppderive{\HMDQ}{22}(\lf{\phisDQ})$ and $\ppderive{\HMDQ}{12}(\lf{\phisDQ})$.

Since the flux linkage~$\lf{\phisDQ}$ is unknown, what is actually obtained is $H_{11}(\lf{\isDQ})$, $H_{12}(\lf{\isDQ})$ and $H_{22}(\lf{\isDQ})$, where $H_{kl}(\isDQ):=\ppderive{\HMDQ}{kl}\bigl(\PhisDQ(\isDQ)\bigr)$ and $\PhisDQ$ is the inverse of the flux-current relation~\eqref{eqn:model:algebraic}.

\subsection{Obtaining the current-flux relations from the $H_{kl}(\isDQ)$}\label{sec:proposed:method}
By the very definition of~$\PhisDQ$, we can write $\isDQ=\nabla\HMDQ\bigl(\PhisDQ(\isDQ)\bigr)$, or, more explicitly,
\begin{IEEEeqnarray*}{rCl}
	\isD &=& \pderive{\HMDQ}{1}\bigl(\PhisD(\isD,\isQ),\PhisQ(\isD,\isQ)\bigr)\\
	\isQ &=& \pderive{\HMDQ}{2}\bigl(\PhisD(\isD,\isQ),\PhisQ(\isD,\isQ)\bigr).
\end{IEEEeqnarray*}
Differentiating these relations with respect to $\isD$ and $\isQ$, we find that
\begin{IEEEeqnarray*}{rCl}
	\begin{pmatrix}
		H_{11}(\isDQ) & H_{12}(\isDQ)\\
		H_{12}(\isDQ) & H_{22}(\isDQ)
	\end{pmatrix}
	\begin{pmatrix}
		\pderive{\PhisD}{1}(\isDQ) & \pderive{\PhisD}{2}(\isDQ)\\
		\pderive{\PhisQ}{1}(\isDQ) & \pderive{\PhisQ}{2}(\isDQ)
	\end{pmatrix}
\end{IEEEeqnarray*}
is the identity matrix, which implies
\begin{IEEEeqnarray*}{rCl}
	\begin{pmatrix}
	\pderive{\PhisD}{1}(\isDQ) & \pderive{\PhisD}{2}(\isDQ)\\
	\pderive{\PhisQ}{1}(\isDQ) & \pderive{\PhisQ}{2}(\isDQ)
	\end{pmatrix}&=&
	\begin{pmatrix}
		H_{kl}(\isDQ) & H_{kl}(\isDQ)\\
		H_{kl}(\isDQ) & H_{kl}(\isDQ)
	\end{pmatrix}^{-1}\\
	&=:&\begin{pmatrix}
	L_{dd}(\isDQ) & L_{dq}(\isDQ) \\
	L_{dq}(\isDQ)  & L_{qq}(\isDQ)
	\end{pmatrix}.
\end{IEEEeqnarray*}
We thus know the partial derivatives of $\PhisDQ$, or, equivalently, the integrable differential forms
\begin{IEEEeqnarray*}{rCl}
	d\PhisD(\isDQ) &=& L_{dd}(\isDQ)d\isD + L_{dq}(\isDQ)d\isQ\\
	d\PhisQ(\isDQ) &=& L_{dq}(\isDQ)d\isD + L_{qq}(\isDQ)d\isQ.
\end{IEEEeqnarray*}
To recover $\PhisDQ$, we integrate these forms on paths $\kappa\mapsto\isDQ=\jsDQ(\kappa)$ in the current plane. This yields
\begin{IEEEeqnarray}{rCl}
	\IEEEeqnarraymulticol{3}{l}{\PhisD\bigl(\jsDQ(\kappa)\bigr)-\PhisD\bigl(\jsDQ(\kappa_0)\bigr)
		=\int_{\kappa_0}^\kappa\frac{d\PhisD}{d\zeta}\bigl(\jsDQ(\zeta)\bigr)d\zeta}\nonumber\\* 
	\qquad&=& \int_{\kappa_0}^\kappa L_{dd}\bigl(\jsDQ(\zeta)\bigr)\frac{d\jsD}{d\zeta}(\zeta)d\zeta\nonumber\\
	&&\quad+\>\int_{\kappa_0}^\kappa L_{dq}\bigl(\jsDQ(\zeta)\bigr)\frac{d\jsQ}{d\zeta}(\zeta)d\zeta,\label{eqn:proposal:integral}
\end{IEEEeqnarray}
and a similar expression for~$\PhisQ\bigl(\jsDQ(\kappa)\bigr)$.
Therefore, for each~$\kappa$, a couple $\bigl(\jsDQ(\kappa),\PhisD\bigl(\jsDQ(\kappa)\bigr)\bigr)$ is obtained; provided the current paths $\kappa\mapsto\jsDQ(\kappa)$ cover a sufficient area of the current plane, this yields the desired current-flux relations. The initial value $\PhisDQ(\jsDQ(\kappa_a)\bigr)$ is chosen so as the flux linkage is zero when the current is zero.

Notice the method is completely immune to errors in the knowledge of the stator resistance~$R_s$, since it never explicitly uses its value. Moreover, it is less sensitive than the classical method to imperfections of the power inverter because the base potential $\lf{\vsDQ}$ does not need to be known, whereas the injected potential $\hf{\vsDQ}$ can be fairly large.

\subsection{Experimental results}\label{sec:proposed:xp}
\begin{figure}
	\centering
	\includegraphics{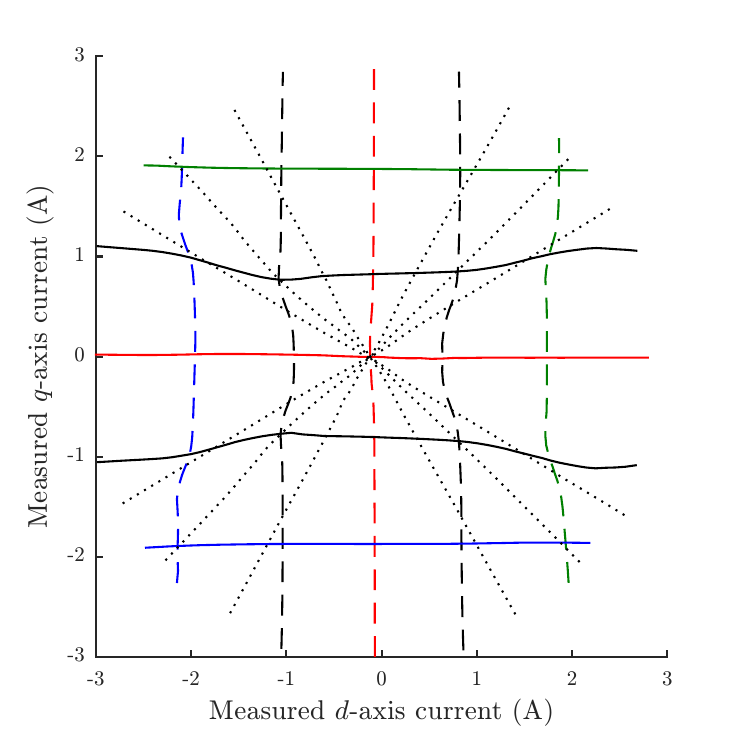}
	\caption{Actual $\lf{\isDQ}$ paths.}
	\label{fig:proposalxp:current-traj}
\end{figure}

The proposed approach was applied with the same experimental conditions as in section~\ref{sec:classical:xp}, with similar desired currents paths. 
Nevertheless, the currents were not enforced by controllers not to disturb the signal injection. As a consequence the actual currents paths slightly differ from the desired ones, see~fig.~\ref{fig:proposalxp:current-traj}; this is not a problem since the true values of~$\lf{\isDQ}$ are known.
The current paths were discretized with steps $\Delta\lf{\isDQ}$ of about~\SI{0.1}{\ampere}; for each current point $\lf{\isDQ}$, the injected signal was a \SI{5}{\second} long square signal of frequency~$\Omega:=\SI{500}{\hertz}$ and amplitude $\hf{\us}:=\SI{40}{\volt}$, slowly rotating at frequency $\foinj:=\SI{1}{\hertz}$,
\begin{IEEEeqnarray*}{rCl}
\hf{\usDQ} &:=& \hf{\us}\begin{pmatrix}
	\cos(2\pi \foinj t) \\
	\sin(2\pi \foinj t)
\end{pmatrix}.
\end{IEEEeqnarray*}
The corresponding current ripple, namely
\begin{IEEEeqnarray*}{rCl}
	\hf{\isDQ} &=& \frac{\hf{\us}}{\Omega}\begin{pmatrix}
		H_{11}(\lf{\isDQ})\cos(2\pi\foinj t) + H_{12}(\lf{\isDQ})\sin(2\pi\foinj t)\\
		H_{12}(\lf{\isDQ})\cos(2\pi\foinj t) + H_{22}(\lf{\isDQ})\sin(2\pi\foinj t) 
	\end{pmatrix},
\end{IEEEeqnarray*}
then directly yields $H_{11}(\lf{\isDQ})$, $H_{12}(\lf{\isDQ})$ and $H_{22}(\lf{\isDQ})$. The interest of slowly rotating the injected signal is to provide many measurements points $\hf{\isDQ}$, hence an accurate determination of $H_{11}(\lf{\isDQ})$, $H_{12}(\lf{\isDQ})$ and $H_{22}(\lf{\isDQ})$.

\begin{figure}
	\centering
	\subfigure[$D$-axis flux linkage in function of $\isDQ$\label{fig:proposalxp:results:d}]{\includegraphics{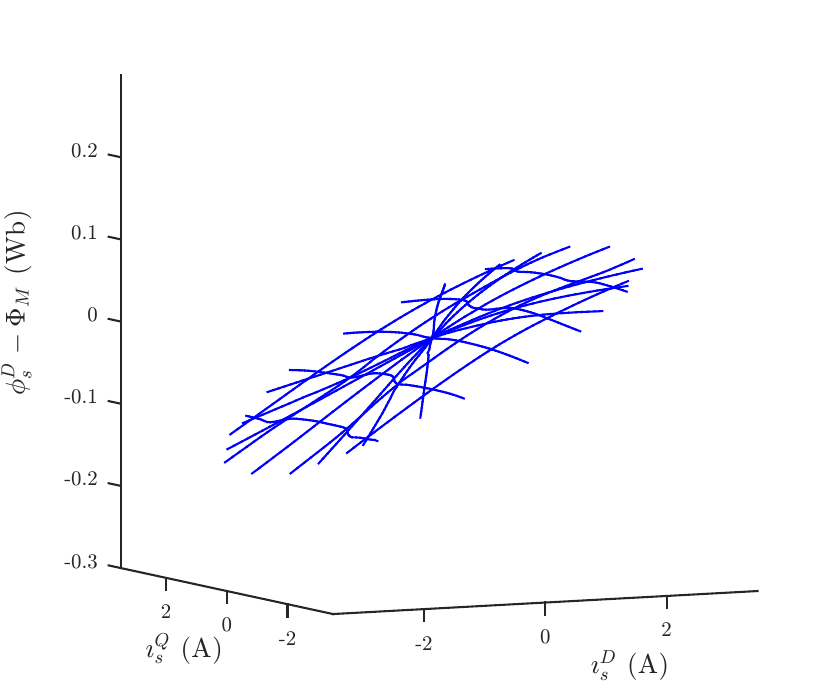}}
	\subfigure[$Q$-axis flux linkage in function of $\isDQ$\label{fig:proposalxp:results:q}]{\includegraphics{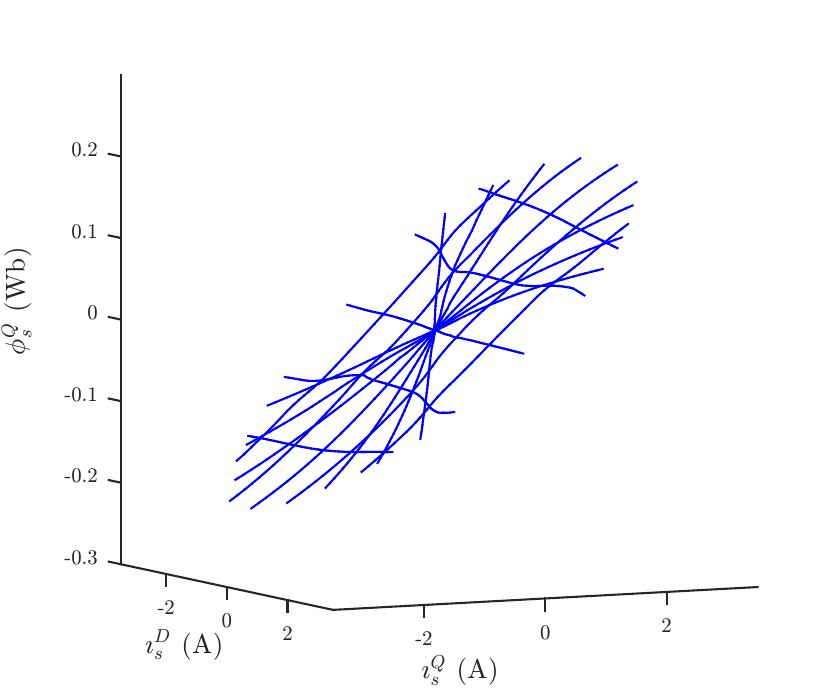}}
	\caption{Current-flux relations obtained by the proposed method.}
	\label{fig:proposalxp:results}
\end{figure}

The flux linkage along a current path is then obtained by discretizing the integrals in~\eqref{eqn:proposal:integral},
\begin{IEEEeqnarray*}{rCl}
	\IEEEeqnarraymulticol{3}{l}{\PhisD\bigl(\jsDQ(\kappa)\bigr)-\PhisD\bigl(\jsDQ(\kappa_0)\bigr)}\\* 
	\quad&\approx& \sum_{\zeta=\kappa_0+1}^\kappa L_{dd}\bigl(\jsDQ(\zeta)\bigr)\Delta\jsD(\zeta)
	+ L_{dq}\bigl(\jsDQ(\zeta)\bigr)\Delta\jsQ(\zeta),
\end{IEEEeqnarray*}
where $\Delta\jsDQ(\zeta):=\jsDQ(\zeta)-\jsDQ(\zeta-1)$. The current-flux curves so obtained are displayed in~fig.~\ref{fig:proposalxp:results}. The consistency can be checked by computing the flux linkage difference at points where the current paths intersect, which should be zero in theory: the maximum relative error is \SI{1.3}{\percent} for the $D$-axis flux and \SI{2.9}{\percent} for the $Q$-axis. Moreover, as with the classical method, $\phisD$ is even with respect to $\isQ$, and $\phisQ$ odd with respect to $\isQ$.

\begin{figure}
	\centering
	\subfigure[$L{dd}$ as a function of~$\phisDQ$\label{fig:proposalxp:hessian:d}]{\includegraphics{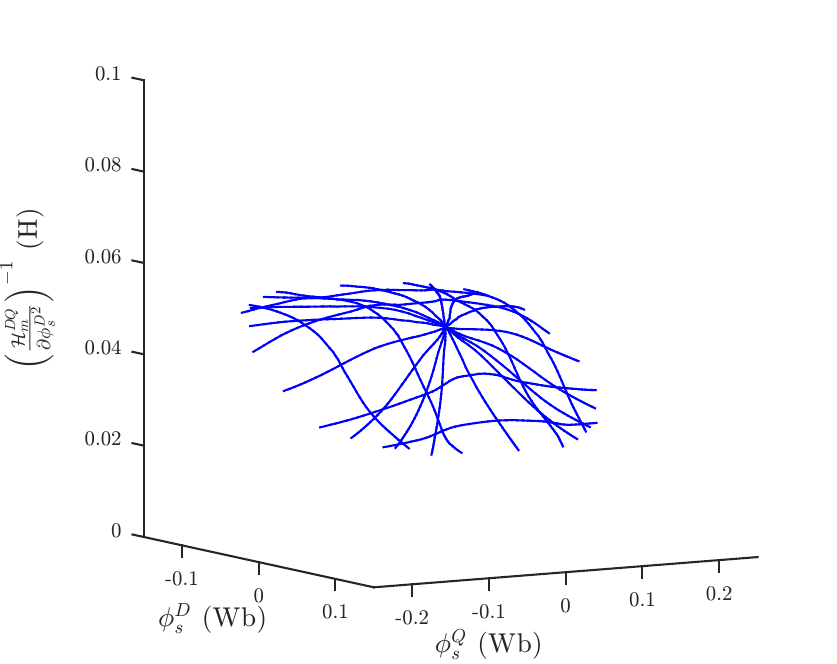}}
	\subfigure[$L{qq}$ as a function of~$\phisDQ$\label{fig:proposalxp:hessian:q}]{\includegraphics{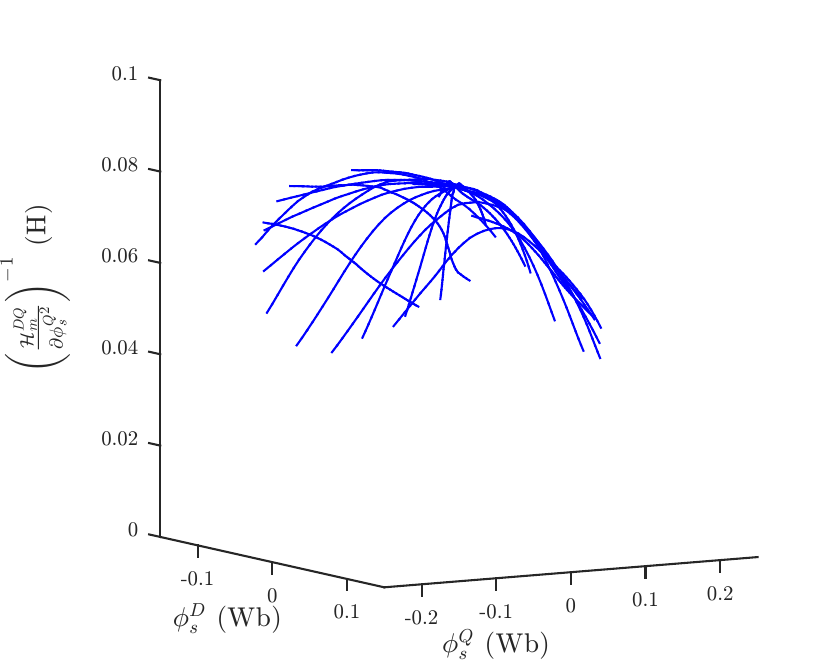}}
	\subfigure[$L{qq}$ as a function of~$\phisDQ$\label{fig:proposalxp:hessian:x}]{\includegraphics{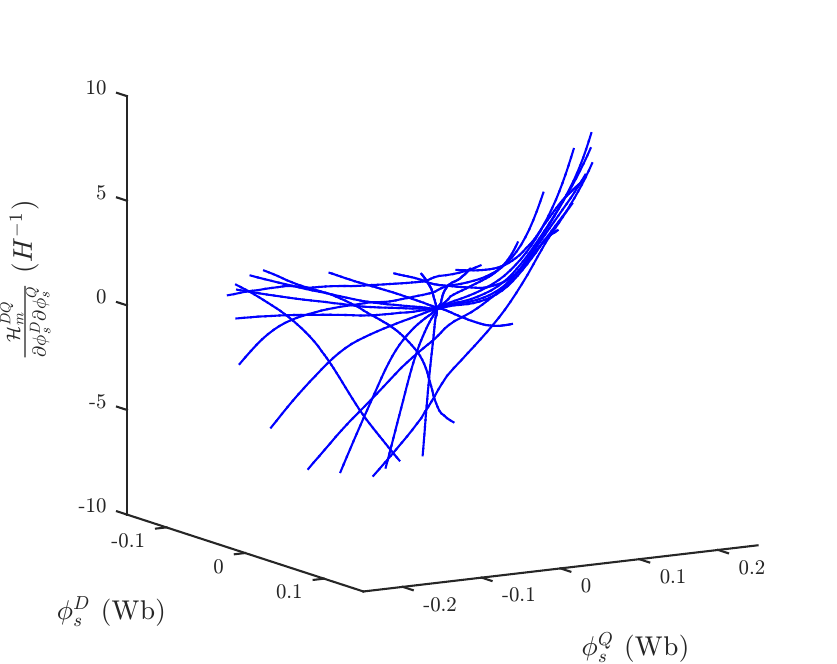}}
	\caption{Coefficients of inductance matrix obtained by the proposed method.}
	\label{fig:proposalxp:hessian}
\end{figure}

Notice the method provides the Hessian matrix of the energy function
\begin{IEEEeqnarray*}{C}
	\begin{pmatrix}\ppderive{\HMDQ}{11} & \ppderive{\HMDQ}{11} \\ \ppderive{\HMDQ}{11}  & \ppderive{\HMDQ}{11}\end{pmatrix},
\end{IEEEeqnarray*}
in function of $\isDQ$ or $\phisDQ$, i.e., the partial derivatives of the current-flux relations. It is a useful piece of information for fitting a parametric model to the current-flux relations; indeed, it is numerically better to fit the derivatives of a function than the function itself. It also provides its inverse matrix,
\begin{IEEEeqnarray*}{C}
	\begin{pmatrix}L_{dd} & L_{dq} \\ L_{dq} & L_{qq}\end{pmatrix},
\end{IEEEeqnarray*}
which is the inductance matrix. As expected from~\eqref{eqn:model:symmetries}, which implies
\begin{IEEEeqnarray*}{rCl}
	\ppderive{\HMDQ}{11}(\phisD,-\phisQ) &=& \ppderive{\HMDQ}{11}(\phisD,\phisQ)\\
	\ppderive{\HMDQ}{22}(\phisD,-\phisQ) &=& \ppderive{\HMDQ}{22}(\phisD,\phisQ)\\
	\ppderive{\HMDQ}{12}(\phisD,-\phisQ) &=& -\ppderive{\HMDQ}{12}(\phisD,\phisQ),
\end{IEEEeqnarray*}
the parity/imparity relations are experimentally satisfied, both for the Hessian matrix and its inverse. This is illustrated for the inductance matrix in in fig.~\ref{fig:proposalxp:hessian}; the (cross)-saturation is even more visible than in fig.~\ref{fig:proposalxp:results}.

\subsection{Comparison with the classical method}\label{sec:proposed:comp}
\begin{figure}
	\centering
	\subfigure[$\isD\mapsto\PhisD(\isD,\isQ)-\PhiM$ for $\isQ=\SI{0}{\ampere}$ (\textcolor{red}{---}), $\isQ=\SI{-2}{\ampere}$ (\textcolor{blue}{---}), $\isQ=\SI{2}{\ampere}$ (\textcolor{green!50!black}{---})\label{fig:comparison:d}]{\includegraphics{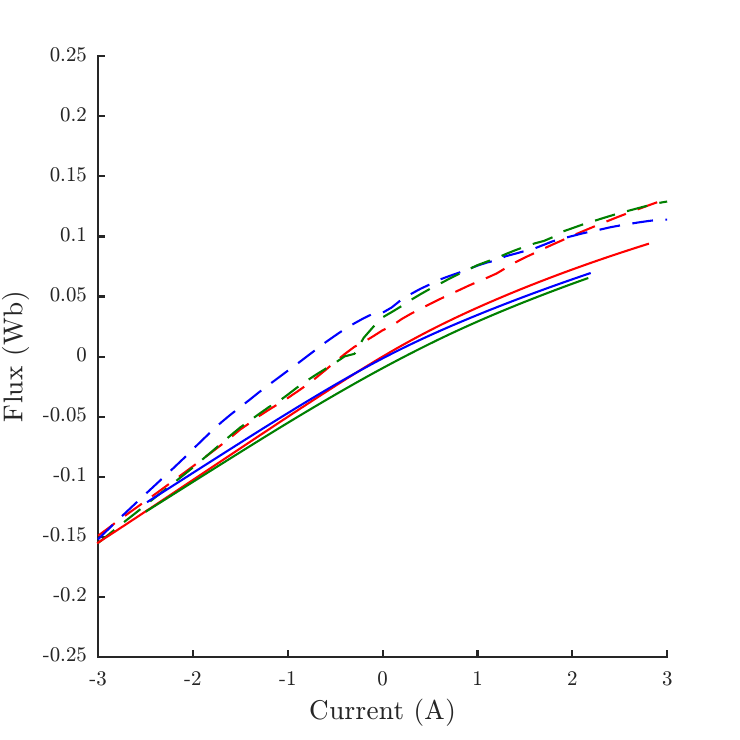}}
	\subfigure[$\isQ\mapsto\PhisQ(\isD,\isQ)$ for $\isD=\SI{0}{\ampere}$ (\textcolor{red}{---}), $\isD=\SI{-2}{\ampere}$ (\textcolor{blue}{---}), $\isD=\SI{2}{\ampere}$ (\textcolor{green!50!black}{---})\label{fig:comparison:q}]{\includegraphics{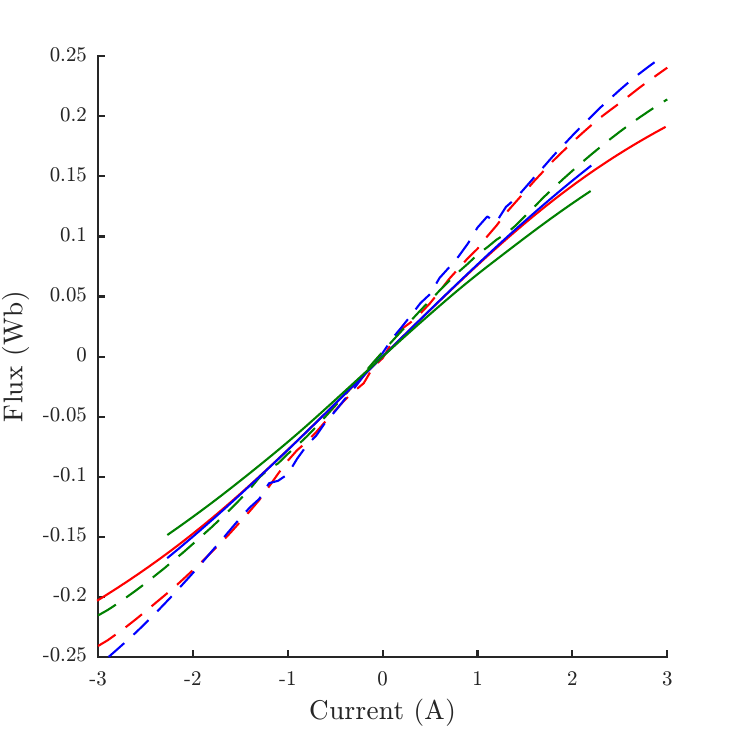}}
	\caption{Comparison of current-flux curves obtained with the classical (dashed lines) and proposed method (solid lines).}
	\label{fig:comparison}
\end{figure}

Fig.~\ref{fig:comparison} compares some current-flux curves obtained by the classical and the proposed method. The two methods yield similar curves, with nevertheless some small differences. Some of the differences could be explained by experimental errors, especially biases in the classical method. Another possibility is that, owing to hysteresis, the flux linkage computed by signal injection is systematically smaller than the flux linkage computed by the classical method; this was noticed for the Synchronous Reluctance Motor in~\cite{CombeMMR2011IEMDC}, and raises the question of which model should be used for sensorless control at low velocity.

\section{Conclusion}\label{sec:conclusion}
We have proposed a method based on signal injection to obtain the saturated current-flux relations of a PMSM from locked-rotor experiments. With respect to the classical method based on time integration, it has the main advantage of being completely independent of the stator resistance; moreover, it is less sensitive to voltage biases due to the power inverter, as the injected signal may be fairly large. 
Besides, the method provides the inductance matrix (as a function of the current or the flux linkage), which is an interesting piece of information by itself, and can also be used to fit a parametric model to the current-flux relations; indeed, it is numerically better to fit the derivatives of a function than the function itself.

	
	\newpage
\bibliographystyle{phmIEEEtran}
\bibliography{biblio}

\end{document}